\documentclass[]{spie}  

\usepackage{amsmath,amsfonts,amssymb}
\usepackage{graphicx}
\usepackage{subfig}
\usepackage[colorlinks=true, allcolors=blue]{hyperref}
\usepackage{siunitx}
\usepackage{float}
\DeclareSIUnit\angstrom{\text{Å}}
\DeclareSIUnit\photon{\text{ph}}
\DeclareSIUnit\arcsec{\text{arcsec}}
\usepackage{amssymb}
\usepackage{xcolor}
\usepackage{url}

\title{End-to-End simulation framework for astronomical spectrographs: SOXS, CUBES and ANDES}

\author[1,23]{A. Scaudo}
\author[1]{M. Genoni}
\author[2]{G. Li Causi}
\author[1]{L. Cabona}
\author[1,21]{M. Landoni}
\author[1]{S. Campana}
\author[3]{P. Schipani}
\author[4]{R. Claudi}
\author[1]{M. Aliverti}
\author[4]{A. Baruffolo}
\author[5]{S. Ben-Ami}
\author[4]{F. Biondi}
\author[3]{G. Capasso}
\author[7]{R. Cosentino}
\author[8]{F. D'Alessio}
\author[1]{P. D'Avanzo}
\author[5]{O. Hershko}
\author[9,10]{H. Kuncarayakti}
\author[11]{M. Munari}
\author[4]{K. Radhakrishnan Santhakumari}
\author[12,13]{G. Pignata}
\author[18]{A. Rubin} 
\author[11]{S. Scuderi} 
\author[8]{F. Vitali}
\author[14]{D. Young}
\author[15]{J. Achrén}
\author[33]{J. A. Araiza-Duran}
\author[17]{I. Arcavi}
\author[4]{F. Battaini}
\author[33]{A. Brucalassi} 
\author[5]{R. Bruch}
\author[4]{E. Cappellaro}
\author[3]{M. Colapietro}
\author[3]{M. Della Valle}
\author[4]{M. De Pascale}
\author[11]{R. Di Benedetto}
\author[3]{S. D'Orsi}
\author[5]{A. Gal-Yam}
\author[7]{M. Hernandez}
\author[9,10]{J. Kotilainen}
\author[3]{L. Marty}
\author[10]{S. Mattila}
\author[5]{M. Rappaport}
\author[4]{D. Ricci}
\author[1]{M. Riva}
\author[4]{B. Salasnich}
\author[14]{S. Smartt}
\author[11]{R. Zanmar Sanchez}
\author[19]{M. Stritzinger}
\author[7]{H. Ventura}
\author[22, 24]{G. Cupani}
\author[22]{M. Porru}
\author[22]{M. Franchini}
\author[22]{R. Cirami}
\author[22]{G. Calderone}
\author[1]{S. Covino}
\author[25]{R. Smiljanic}
\author[26]{M. Monteiro}
\author[4]{A. Balestra}
\author[4]{R. Sordo}
\author[22]{E. Mason}
\author[34]{S. Rousseu}
\author[28]{I. de Castro Leão}
\author[1]{A. Zanutta}
\author[28]{A. de Meideros Martins}
\author[29]{D. Sosnowska}
\author[30]{T. Marquart}
\author[31]{I. Boisse}
\author[26]{S. Sousa}
\author[32]{J. Hernandez}
\author[30]{N. Piskunov}
\author[30]{J. Puschnig}
\author[33]{N. Senna}
\author[28]{B. Martins}
\author[22]{P. Di Marcantonio}
\author[33] {A. Marconi}

\affil[1]{INAF– Osservatorio Astronomico di Brera-Merate, via E. Bianchi 46, I-23807 Merate (LC), Italy;}
\affil[2]{INAF– Istituto di Astrofisica e Planetologia Spaziali, via Fosso del Cavaliere 100, Roma – Italy;}
\affil[3]{INAF – Osservatorio Astronomico di Capodimonte, Sal. Moiariello 16, I-80131, Naples, Italy;}
\affil[4]{INAF – Osservatorio Astronomico di Padova, Vicolo dell’Osservatorio 5, I-35122, Padua, Italy;}
\affil[5]{Weizmann Institute of Science, Herzl St 234, Rehovot, 7610001, Israel;}
\affil[6]{Max-Planck-Institut für Extraterrestrische Physik, Giessenbachstr. 1, D-85748 Garching, Germany;}
\affil[7]{FGG-INAF, TNG, Rambla J.A. Fernández Pérez 7, E-38712 Breña Baja (TF), Spain;}
\affil[8]{INAF – Osservatorio Astronomico di Roma, Via Frascati 33, I-00078 M. Porzio Catone, Italy;}
\affil[9]{Finnish Centre for Astronomy with ESO (FINCA), FI-20014 University of Turku, Finland;}
\affil[10]{Tuorla Observatory, Dept. of Physics and Astronomy, FI-20014 University of Turku, Finland;}
\affil[11]{INAF – Osservatorio Astroﬁsico di Catania, Via S. Soﬁa 78 30, I-95123 Catania, Italy;}
\affil[12]{Instituto de Alta Investigación, Universidad de Tarapacá, Casilla 7D, Arica, Chile;}
\affil[13]{Millennium Institute of Astrophysics (MAS), Santiago, Chile;}
\affil[14]{Astrophysics Research Centre, Queen’s University Belfast, Belfast, BT7 1NN, UK;}
\affil[15]{Incident Angle Oy, Capsiankatu 4 A 29, FI-20320 Turku, Finland;}
\affil[16]{Centro de Investigaciones en Optica A. C., 37150 León, Mexico;}
\affil[17]{Tel Aviv University, Department of Astrophysics, 69978 Tel Aviv, Israel;}
\affil[18]{ESO, Karl Schwarzschild Strasse 2, D-85748, Garching bei München, Germany;}
\affil[19]{Aarhus University, Ny Munkegade 120, D-8000 Aarhus, Denmark;}
\affil[20]{IFPU - Institute for Fundamental Physics of the Universe, via Beirut 2, 34151 Trieste, Italy;}
\affil[21]{INAF– Osservatorio Astronomico di Cagliari, via della Scienza 5, 09047, Selargius (CA), Italy;}
\affil[22]{INAF - Osservatorio Astronomico di Trieste, via G. B. Tiepolo 11, 34131 Trieste, Italy;}
\affil[23]{University of Insubria, Via J.H. Dunant, 3, 21100 Varese, Italy;}
\affil[24]{IFPU - Institute for Fundamental Physics of the Universe, via Beirut 2, 34151 Trieste, Italy;}
\affil[25]{Nicolaus Copernicus Astronomical Ctr., ul. Bartycka 18, 00-716 Warszawa, Poland;}
\affil[26]{Instituto de Astrofisica e Ciencias do Espaco, Universidade do Porto, CAUP, Rua das Estrelas, 4150-762 Porto Portugal;}
\affil[27]{French National Centre for Scientific Research - CNRS;}
\affil[28]{Univ. Federal do Rio Grande do Norte, Rio Grande do Norte, Brazil;}
\affil[29]{Observatoire Astronomique de l’Universite de Geneve, Chemin Pegasi 51, Versoix, CH-1290, Switzerland;}
\affil[30]{Uppsala University, Department of Physics and Astronomy; Astronomy and Space Physics, Box 516, 751 20 Uppsala, Sweden;}
\affil[31]{Lab. d'Astrophysique de Marseille, Marseille, France;}
\affil[32]{Instituto de Astrofísica de Canarias, San Cristobal de la Laguna, Spain;}
\affil[33]{INAF - Osservatorio Astrofisico di Arcetri, Largo E. Fermi 5, I-50125 Firenze, Italy;}
\affil[34]{Observatoire de la Côte d'Azur - OCA, France;}

\begin{document} 
\maketitle

\begin{abstract}
We present our numerical simulation approach for the End-to-End (E2E) model applied to various astronomical spectrographs, such as SOXS (ESO-NTT), CUBES (ESO-VLT), and ANDES (ESO-ELT), covering multiple wavelength regions. The E2E model aim at simulating the expected astronomical observations starting from the radiation of the scientific sources (or calibration sources) up to the raw-frame data produced by the detectors. The comprehensive description includes E2E architecture, computational models, and tools for rendering the simulated frames. Collaboration with Data Reduction Software (DRS) teams is discussed, along with efforts to meet instrument requirements. The contribution to the cross-correlation algorithm for the Active Flexure Compensation (AFC) system of CUBES is detailed.
\end{abstract}

\section{INTRODUCTION}
End-to-End (E2E) Simulators are valuable tools in the development of astronomical instrumentation. These simulators are used by multiple consortiums at various stages of instrument development, providing critical insights and feedback throughout the entire development cycle. Additionally, during operation, they serve as reliable tools to aid in setting calibration procedures and planning observations. The primary objective of the E2E simulator is to generate realistic synthetic raw frames by simulating the entire observation chain. This process begins with the radiation of scientific or calibration sources and extends to the final raw-frame data produced by the detectors. This paper presents our simulation approach and details how we exploit the capabilities of the E2E simulator for different astronomical spectrographs. Specifically, we focus on three instruments: SOXS (ESO-NTT), CUBES (ESO-VLT), and ANDES (ESO-ELT), which cover multiple wavelength regions and are at different phases of their respective development cycles.

For the SOXS instrument, the E2E simulator has been employed during the Assembly, Integration, and Testing (AIT) phase, providing valuable feedback to ensure the instrument meets its design specifications. In the case of the CUBES instrument, the E2E simulator is playing a crucial role in developing the prototype Data Reduction Software (pDRS) and testing the cross-correlation algorithm of the Active Flexure Compensation (AFC) system. Additionally, the simulation framework Pyechelle has been employed to provide synthetic raw frames during the early design phase of the ANDES Spectrograph.

This paper is organized as follows: Section \ref{sec:e2e} presents the E2E Simulator, detailing its architecture and the computational models and framework used for rendering the simulated frames, such as Pyechelle and Pyxel. Sections \ref{sec:soxs_instrument}, \ref{sec:cubes_instrument}, and \ref{sec:andes_instrument} discuss the SOXS, CUBES, and ANDES instruments, respectively, and how the E2E simulator has been applied to each of these instruments.
\label{sec:intro}  

\section{Simulation Approaches}
Two different approaches have been employed to provide simulation frames. For the SOXS and CUBES spectrographs, we developed our own numerical simulator customized to the specific needs and characteristics of each instrument. The E2E simulator is presented in detail in the next Subsection \ref{sec:e2e_architecture}, where we discuss its modular architecture and functionalities. Meanwhile, for the ANDES spectrograph, we opted to use the Pyechelle framework (see Subsection \ref{sec:pyechelle}). Pyechelle offers a robust and flexible simulation environment that aligns well with the early design phase needs of ANDES.
\label{sec:e2e}
\begin{figure} [htp]
\begin{center}
\begin{tabular}{c} 
\includegraphics[width=0.75\linewidth]{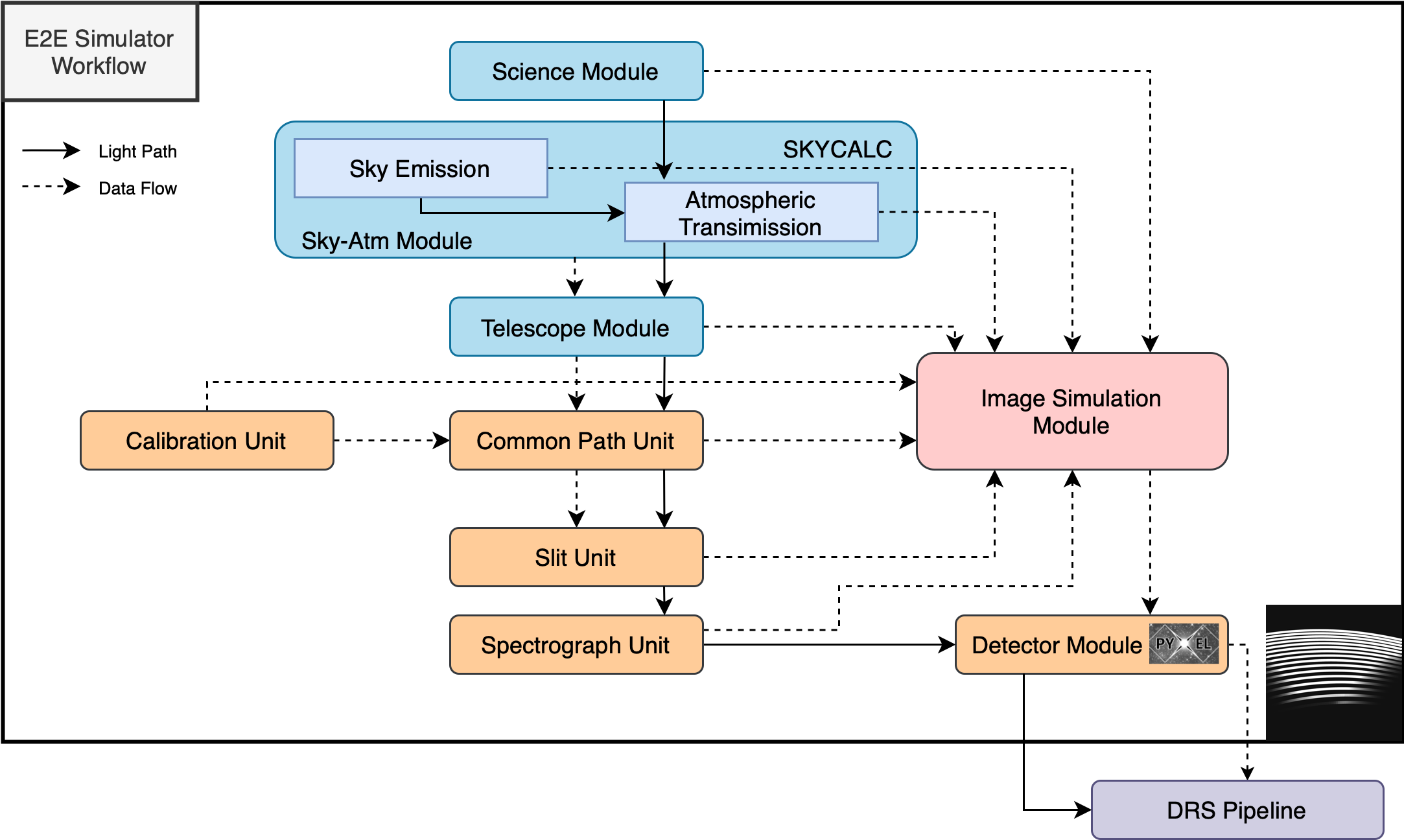}
\end{tabular}
\end{center}
\caption[example] 
{ \label{fig:e2e_arc} 
E2E simulator workflow schematic description. The solid arrows represent the real light path, while the dashed arrows show: the simulation data-flow, how the different modules and units are interfaced and their connection to the simulator core, which is the Image Simulation Module. Orange blocks are units of the Instrument Module, while blue blocks are related to simulation modules independent from the specific instrument.}
\end{figure} 
\subsection{End-to-End Simulator}
\label{sec:e2e_architecture}
The End-to-End simulator architecture is highly modular, composed by different modules (each one with specific tasks and functionalities), units and interfaces, as described in the schematic workflow of Fig. \ref{fig:e2e_arc}. 
Modularity and flexibility are key points in the definition of tasks and interfaces among different modules or units to allow this tool to be scalable and adaptable to different instruments and types of simulations (science versus calibrations frames).
The simulator is written in Python 3.9 and uses specific libraries and APIs for interfacing with other software, like commercial optical ray tracing Zemax-OpticStudio®.
In the following a brief description of the modules and units showed in the workflow diagram is given; more details can be found here \cite{soxs_gen_2022_e2e}.

\subsubsection{Science Module}
\label{sec:e2e_architecture_sci_module}
\begin{figure} [ht]
\begin{center}
\begin{tabular}{c} 
\includegraphics[width=0.9\linewidth]{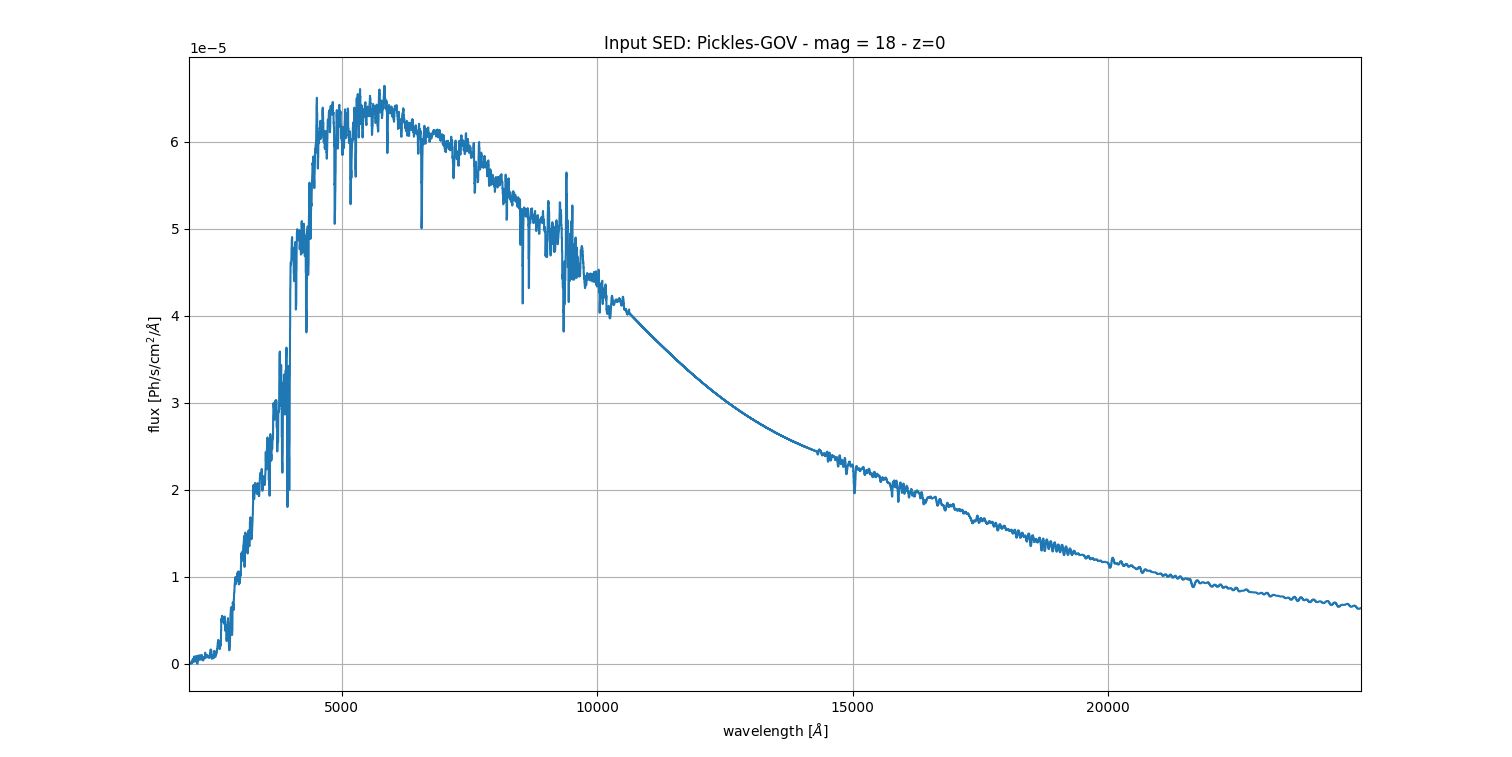}
\end{tabular}
\end{center}
\caption[example] 
{ \label{fig:soxs_sci_module} 
Example of a synthetic 1D spectrum for a G0V star (Pickels model) sampled at resolution of 0.5 \AA, mag 18 (V band of Johnson-Cousins filters, in Vega magnitude system) and at $z=0$.}
\end{figure}

The Science Module has the purpose to generate a synthetic 1D spectrum related to a specific astronomical source, at a resolution higher than the one selected for the instrument simulation. The 1D spectral flux density distribution is obtained by a set of parameters (e.g. object type - spectral type for stars -, magnitude and redshift) or by loading the spectrum from a user-defined library. 
An example of a synthetic 1D spectrum for a G0V star (Pickels model) sampled at resolution of 0.5  Angstrom, is shown in the Fig. \ref{fig:soxs_sci_module}.

\subsubsection{Sky-Atmosphere Module}
\label{sec:e2e_architecture_sky_module}
The task of this module is to model the scattering, absorption and emission occurring in the Earth's atmosphere. 
This is done by calling sky-emission and atmospheric transmission spectra of a dedicated library built using the ESO SkyCalc tool (available at the web page\cite{skycalc}). 
The sky radiance spectrum loaded is in units of ph/s/m$^2$/$\mu$m/arcsec$^2$, thus it is first calculated for the on-sky area related to the slit length and selected slit width according to the simulation (both in  arcsec) and then the units are converted in ph/(s cm\textsuperscript{2} \AA).
\begin{figure}[htp]
\begin{center}{\label{fig:soxs_sky_rad}\includegraphics[width=0.425\linewidth]{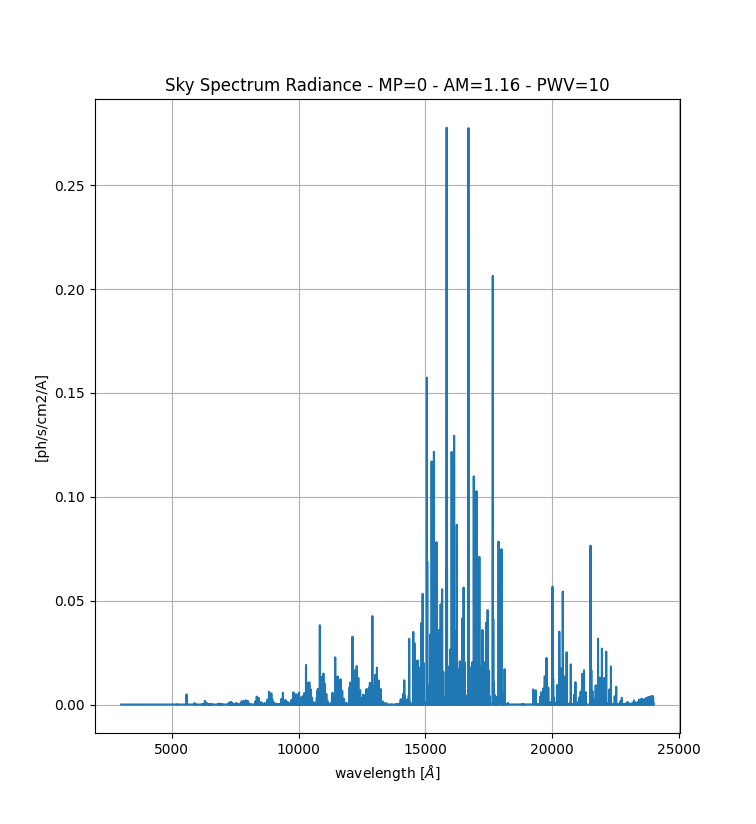}} \qquad
{\label{fig:soxs_sky_tra}\includegraphics[width=0.425\linewidth]{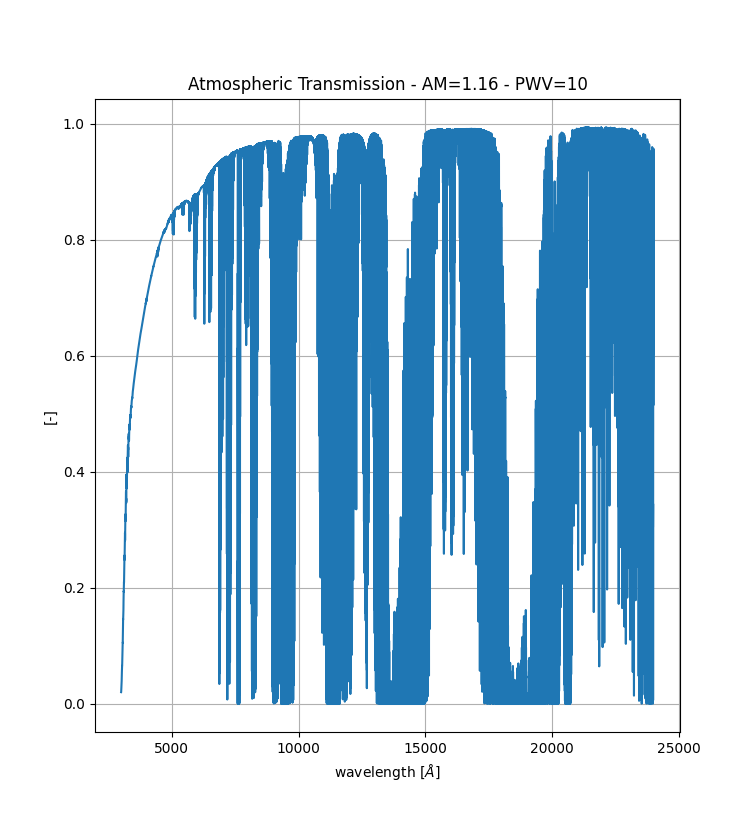}}
\caption[example] 
{ \label{fig:soxs_sky_module}  
Example of a sky radiance spectrum and atmospheric transmission calculated through the ESO-SkyCalc tool. New moon, air-mass=1.16 and precipitable water vapor (PWV) = 10  mm 
}
\end{center}
\end{figure}

\subsubsection{Telescope Module}
The aim of this module is to predict the telescope throughput based on the available telescope mirrors reflectivity data.

\subsection{Instrument Module}
The composition of the instrumental module depends on the specific spectrograph to be simulated and its characteristics. For instance, the calibration unit is responsible for simulating the Spectral Energy Distribution (SED) of the calibration sources. This unit requires information about the types of lamps to be simulated, the instrument's resolution, and the type of calibration mask. The slit unit computes the slit efficiency, i.e. the fraction of light passing through the pseudo-slit reformatted at the spectrograph entrance slit. This is related to the design configuration and/or resolution mode. 
Factors considered in calculating slit efficiency include the instrument's entrance aperture dimensions (width and length), the optical quality (PSF FWHM) up to the slicer focal plane, and the seeing conditions (FWHM at \SI{500}{\nano\meter}). Optionally, the PSF can be convolved with a given spatial profile, which is useful for simulating extended targets. This unit extracts the spectral format data, which can vary depending on the instrument to be simulated, the instrument throughput, and the slice PSF maps from the Spectrograph Optical Files designed in Zemax-OpticStudio®. The accuracy of the aberrations, distortion, and diffraction effects is determined by the parameters used in these specific optical design files. These maps are then used by the Image Simulation Module to render the synthetic frames.
\subsubsection{Image Simulation Module}
\label{sec:e2e_architecture_img_module}
This portion of the simulator is the kernel of the whole system and put together the outputs of all the other modules and units. This piece of software is responsible for rendering the photons distribution of each spectral resolution element for each order as should be detected at the level of spectrographs focal plane by the detectors. 
In particular, the module first interpolates on a sub-pixel scale, of which the oversampling can be set according to the required simulation accuracy, all the instrumental data regarding wavelength, image centroid coordinates, sampling (in both main dispersion and spatial directions), slit image tilt, average PSF map and efficiency. 
Then, it produces the photons distribution of each wavelength in sub-pixel scale by convolving the slit (long slit or pinhole masks according to the specific science or calibration frame type) image with the corresponding PSF map. The spectral slit images are properly sized and tilted according to the sampling and tilt variation along each order, and scaled for the integrated spectral flux and efficiency. 
The code architecture has been developed in order to properly exploit the functionality of {\tt numba}\cite{numba} and to take advantage of the parallel features of the concurrent Python package, such that the different diffraction orders (or a fraction of them) can run in parallel. 
Once all the spectral images have been piled up to sum their photons distribution at sub-pixel scale, the synthetic frame is re-binned to pixel scale. Then, the photon noise and all other specific detector noises are added by the Detector Module.

\subsubsection{Detector Module}
The detector module utilizes the Pyxel simulation framework \cite{pyxel_2022} to model various detector effects on rendered data at the pixel level. This tool operates through a series of cascading modules that simulate the entire detection chain. For instance, shot noise can be modeled at the photon collection level, while quantum efficiency (QE), dark current (DC), and cosmic ray effects are incorporated at the charge generation level. Additional effects such as read-out noise (RON), bias levels, fixed pattern noise (caused by pixel non-uniform response, PRNU), and pixel cross-talk (due to charge diffusion) can be added at different stages of the simulation. This framework also enables the generation of bias and dark frames for calibration purposes.
\subsection{Pyechelle}
Pyechelle (for details see \cite{pyechelle_2019}) was selected as the 'engine' to generate the ANDES simulated images in our E2E simulation cascade. Pyechelle is an open-source Python package designed to simulate realistic raw-frame spectra starting from the Zemax-OpticStudio® models. This solution brought great benefits in terms of resources, as we were able to focus more on implementing the effects we aimed for in the final image. This framework can quickly generate high-fidelity spectra for both astronomical and calibration sources, thanks to its support for Numba and CUDA. Information extracted from the Zemax model by various subsystems is stored in specific .hdf files. These files contain details about the spectrograph characteristics, fiber information, spectral orders, and a set of PSF maps. The simulations are performed by applying wavelength-dependent affine transformation matrices to map the aperture onto the detector plane. A wavelength dependent PSF is then applied by alias sampling on the transformed slit images to properly account for optical aberrations. For detector simulation, we have implemented Pyxel in our simulation cascade, as in the case of our E2E model.
\label{sec:pyechelle}

\section{The SOXS Instrument}
\label{sec:soxs_instrument}
\subsection{SOXS instrument overview}
SOXS is a wide-band spectrograph for the ESO-NTT in La Silla (it will be installed at one of the Nasmyth foci of the NTT) covering in a single exposure the spectral range from the UV to the NIR (350-2000 nanometer). Its central structure (the common path) supports two distinct spectrographs, one operating in the UV-VIS 350-850 nanometer and the other in the NIR 800-2000 nanometer wavelength ranges, respectively. Both spectrographs can operate independently at different resolutions according to the slit widths: R$\approx$10000 with $0.5"$ slit, R$\approx$4500 with $1"$ slit and R$\approx$3300 with $1.5"$ slit. See general overview here \cite{soxs_update, soxs_gen_2022}.


The two arms are fed by the light coming through a common opto-mechanical system (the Common Path \cite{soxs_cp}). It redirects the light from the telescope focus to the spectrograph slits through relay optics reducing the F/number (from F/11 to F/6) and compensating for the atmospheric dispersion (only in the UV-VIS). The Common Path provides also the mechanism to drive the light to/from the other instrument subsystems, i.e. the acquisition camera and the calibration unit.

The Acquisition Camera subsystem has different functions: acquisition of the target for the spectrographs, monitoring of spectrographs co‐alignment and light imaging. It consists of a moving stage with positions enabling the various functions, a folding mirror, a filter wheel ({\it u, g, r, i, z, y} LSST and V Johnson bands) front of a compact camera which relays the Nasmyth focus (field of view of $3.5\times 3.5$ arcmin) on the  13.0 $\mu$m, $1024\times1024$ pixel detector CCD. See for details \cite{acq_guid_soxs}.

\begin{figure} [ht]
\begin{center}
\begin{tabular}{c} 
\includegraphics[width=0.35\linewidth]{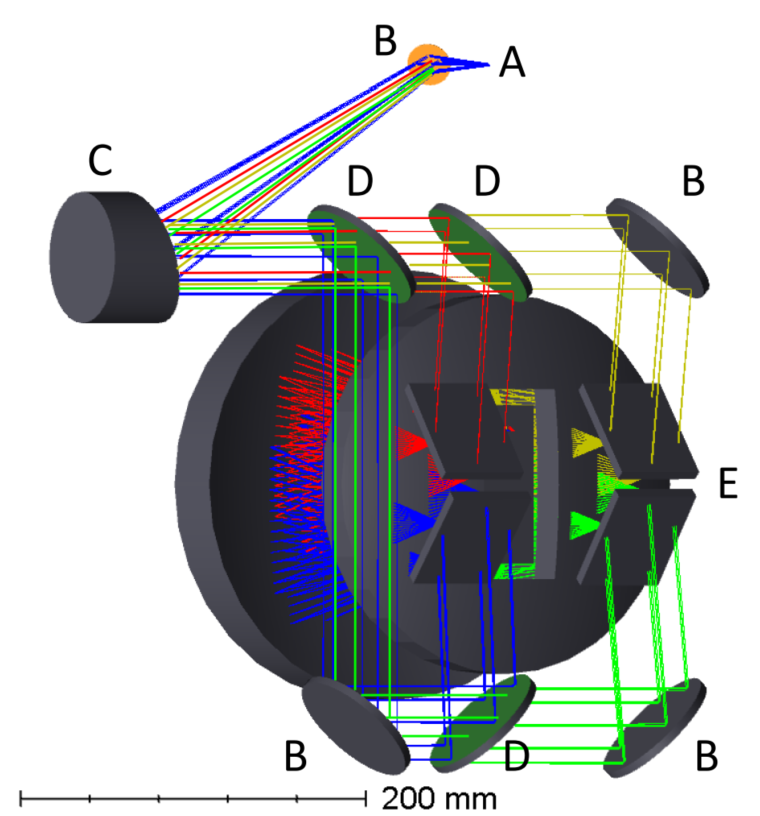}
\end{tabular}
\end{center}
\caption[example] 
{ \label{fig:soxs_vis} 
SOXS UV-VIS spectrograph complete optical layout. A. slit plane, B. reflective mirrors, C. OAP Collimator, D. dichroic filters, E. gratings.}
\end{figure}

\begin{figure} [ht]
\begin{center}
\begin{tabular}{c} 
\includegraphics[width=0.55\linewidth]{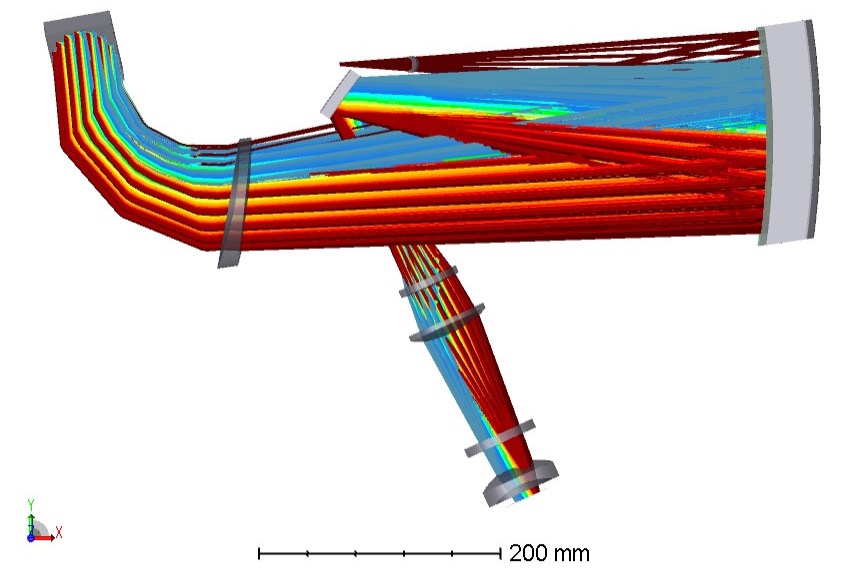}
\end{tabular}
\end{center}
\caption[example] 
{ \label{fig:soxs_nir} 
SOXS NIR spectrograph complete optical layout. The optical beams goes the double pass collimating and cross-dispersing optics. The main disperser is a $44\deg$ blaze angle echelle grating. A field mirror redirect the beam down to the camera.}
\end{figure}

The UV-VIS spectrograph arm is based on a novel multi-grating concept (for details\cite{MITS} and Fig. \ref{fig:soxs_vis}), in which the incoming beam is partitioned into four polychromatic beams using dichroic surfaces, each covering a waveband range of $\sim 100$ nanometer, named as quasi-order. Each quasi-order is diffracted by an ion-etched grating. The detector, located 4 mm behind the camera field flattener back surface, is an e2V CCD44-82 CCD ($2048\times 4096$ pixels, pixel size 15 micrometer, see for details\cite{soxs_vis}).
The near infrared spectrograph, showed in Fig. \ref{fig:soxs_nir}, is a cross-dispersed echelle, with R$\sim 5000$ (for $1''$ slit), covering the wavelength range from 800 to 2000 nanometer with 15 orders (see for details\cite{soxs_optical_design}). It is based on the 4C concept, characterized by a very compact layout, reduced weight of optics and mechanics, and good stiffness. The spectrograph is composed of a double-pass collimator and a refractive camera, an R-1 grating as main disperser and a prism-based cross disperser. The detector is a Teledyne H2RG array operated at 40K, $2048\times 2048$ pixels, pixel size 18 micrometer (see for details\cite{soxs_nir_paper}).
\newline


The calibration unit, provides the calibration spectra to remove the instrument signature. The calibration spectra are generated using a synthetic light source, adopting an integrating sphere equipped with lamps suitable for wavelength and flux calibrations across the full wavelength range of the instrument (350-2000 nanometer); for details see\cite{soxs_cal}. The following lamps are used:
\begin{itemize}
  \item Quartz-tungsten-halogen (QTH) lamp, for flat field frames 500-2000 nanometer;
  \item Deuterium (D2) lamp, for flat field 350-500 nanometer (used simultaneously with QTH lamp for UV-VIS arm);
  \item ThAr hollow cathode lamp, for UV-VIS wavelength calibration;
  \item Ne-Ar-Hg-Xe pen-ray lamps bundled together, for NIR wavelength calibration. The individual lamps are controlled to operate together as one lamp.
\end{itemize}

All the subsystems of the instrument are now under an intense AIT phase at the Astronomical Observatory of Padova, Italy, as detailed in several contributions.\cite{soxs_gen_2024,soxs_cp_2022,soxs_acq_2022,soxs_nir_2022}.
\subsection{Assembly, Integration and Testing Phase of the NIR spectrograph}
\begin{figure}[htp]
\begin{center}
\subfloat[Real]{\label{fig:soxs_real_ne}\includegraphics[width=0.47\linewidth]{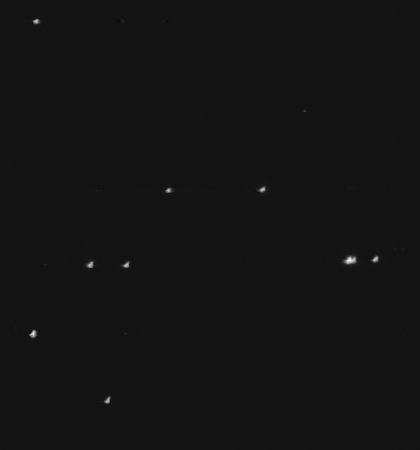}} \qquad
\subfloat[Simulated]{\label{fig:soxs_sim_ne}\includegraphics[width=0.47\linewidth]{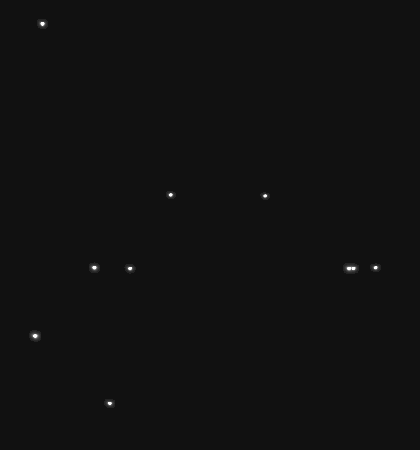}}
\caption[example] 
{ \label{fig:soxs_ne_real_sim}  
Portion of the detector illuminated by a single pinhole with a Xe-lamp, comparing a set of real (\ref{fig:soxs_real_ne}) and simulated (\ref{fig:soxs_sim_ne})  Spectral Resolution Elements (SREs). It is evident from this comparison that the positions of the SREs in the real images correctly match the positions obtained in the simulated images.
}
\end{center}
\end{figure}

For the SOXS instrument, the E2E simulator has been an important tool during the Assembly, Integration, and Testing (AIT) phase. The simulator has provided valuable feedback, ensuring that the instrument meets its design specifications. One of the critical aspects of this phase was to verify the alignment of the spectrograph while comparing it against the simulated images. These simulated frames represent a nominally aligned instrument, serving as a benchmark for the ideal performance of the spectrograph. During the AIT phase, the E2E simulator generated high-fidelity synthetic images that served as a reference for the expected output. Figure \ref{fig:soxs_ne_real_sim} illustrates this process by showing a portion of the detector illuminated by a single pinhole with a Xe-lamp. The figure compares a set of real and simulated Spectral Resolution Elements (SREs). From this comparison, it is evident that the position of the SREs in the real images correctly matches the positions obtained in the simulated images. This initial test was one of the first step towards the alignment process of the spectrograph's optical components.

\begin{figure} [h!]
\begin{center}
\begin{tabular}{c} 
\includegraphics[width=0.95\linewidth]{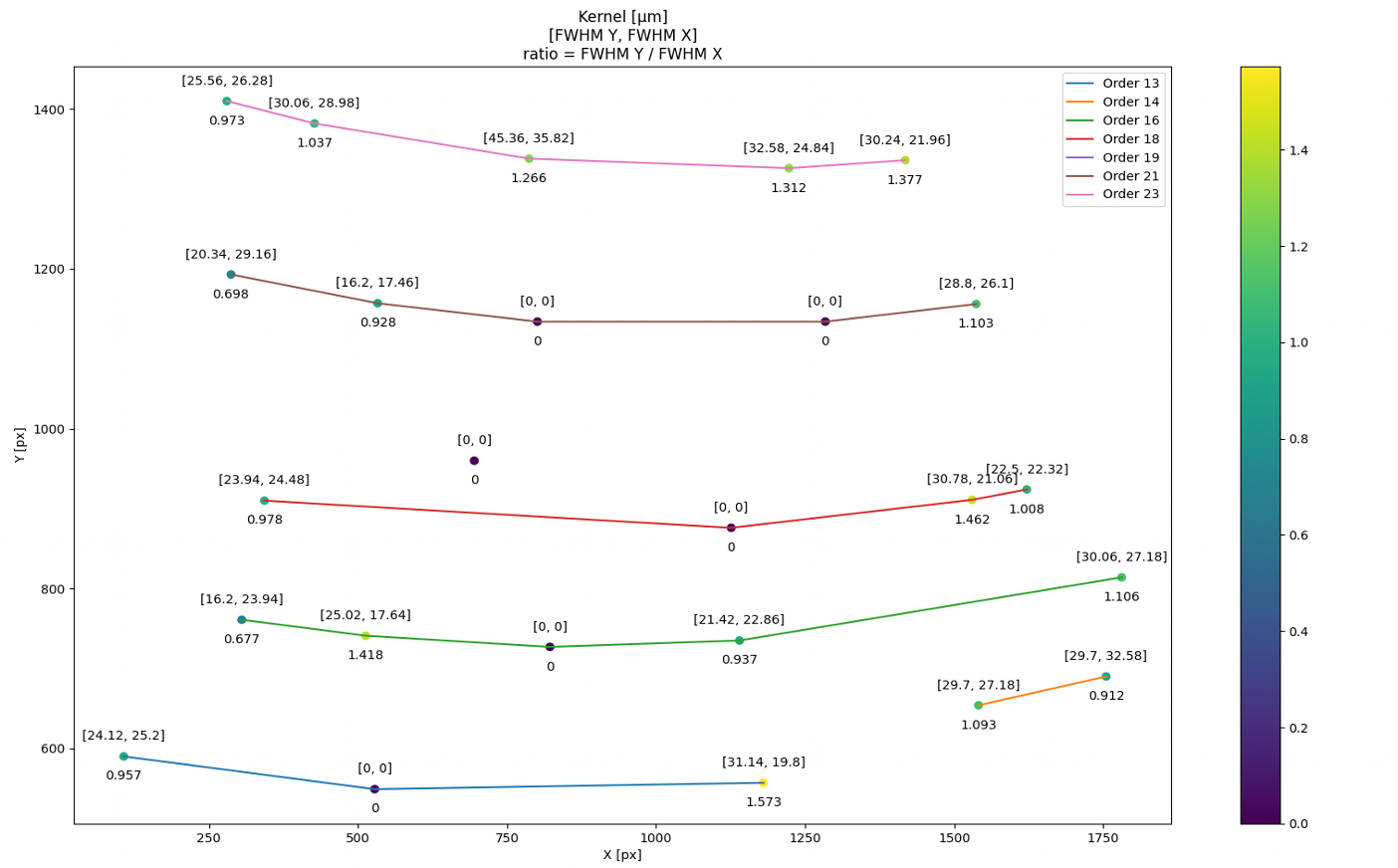}
\end{tabular}
\end{center}
\caption[example] 
{ \label{fig:soxs_kernel} 
Full Width at Half Maximum (FWHM) values in both X and Y directions, and their ratio, across different wavelengths and orders. The kernel values [$\mu$m] represent the difference between the real PSF profile and the simulated one, measured in micrometers. A kernel value of 0 indicates that the quality of the real PSF (FWHM X and Y) is almost equal to the nominal value.
}
\end{figure}

Another relevant test involved the deconvolution of a set of real SREs that were selected to cover the detector evenly (as shown in \cite{soxs_gen_2024}). This process was done to obtain the Point Spread Functions (PSFs) of the different SRE and compare them with the nominal ones used to produce the simulated images. 
Figure \ref{fig:soxs_kernel} represents the Full Width at Half Maximum (FWHM) values in both X and Y directions, as well as the ratio of FWHM Y to FWHM X, across different wavelengths and orders. Specifically: 
\begin{itemize}
    \item Kernel [$\mu$m]: Represents the difference between the real PSF profile and the simulated one, in units of micrometers. A kernel value of 0 indicates that the quality of the real PSF (FWHM X and Y) is almost equal to the nominal value.
    \item Ratio = FWHM Y / FWHM X: Indicates the shape of the PSF, where a ratio close to 1 suggests a more circular PSF, while the opposite indicates elongation.
\end{itemize}
The plotted points show that across the various spectral orders, the PSF size and shape of the real SREs align well with the simulated ones. In general, the image quality is nominal at the center of the orders but slightly degrades towards the edges of the free spectral range. More specifically, in the spectral direction, the difference between the FWHM of the real images with respect to the simulated ones is approximately 1 pixel (18 $\mu$m); while, in the spatial direction, the real blurring is larger (up to 30 $\mu$m, about 1.5 pixels). A portion of this astigmatism is known from the NIR spectrograph nominal design and the alignment was optimized to minimize the aberration of the spectral direction to enhance the instrument resolving power; so this result confirm the expected spectrograph behaviour.

The successful alignment and image quality verification using the E2E simulator underscore its importance in the development of the SOXS instrument, with the iterative process of simulation, testing, and adjustment being highly valuable in achieving the final status of the spectrograph alignment.

\newpage
\section{The CUBES Instrument}
\label{sec:cubes_instrument}
\subsection{CUBES instrument overview}
CUBES will be mounted in one of the Unit Telescope (UT1, UT2 or UT3) at the VLT's Cassegrain focus, and its instrument baseline is composed of several components. These include a front-end unit, which incorporates an acquisition and guiding path. A fore-optics subsystem that encompasses an atmospheric dispersion corrector (ADC), and a re-imaging unit that projects the light to the image slicer. The instrument, in fact, features two image slicers to enable different resolutions. and a calibration unit. Finally, CUBES has two spectrograph arms: arm 1 (blue-optimized) and arm 2 (red-optimized). Both arms are equipped with transmission gratings that have a high groove density, working at the first order, and 9k × 9k scientific detectors.
A detailed description of the CUBES instrument is outside the scope of this paper; see \cite{cubes_2024} for more information. Here, we briefly outline its main characteristics and subsystems.

\subsection{Front-end}
The front-end includes the Acquisition and Guiding (A\&G) subsystem, which handles the initial acquisition of the science target and provides secondary guiding capabilities during exposure.
\subsection{Fore-optics}
The fore-optics subsystem produces a collimated beam for the ADC and magnifying optics, which then supply the image slicer. The image slicer reformats the science object into narrower slits at the spectrograph's input, allowing for higher resolution than a traditional slit. Two image slicers are included, offering both high- and low-resolution modes (HR and LR, respectively).
\subsection{Calibration Unit}
The Calibration Unit provides light sources for various calibration procedures, including flat fielding, wavelength calibration, simultaneous wavelength calibration, alignment, and flexure correction, as an Active Flexure Compensation (AFC) System is required.
\subsection{The spectrograph}
The spectrograph subsystem consists of two arms. Each arm collimates the beam, disperses the light using a transmission grating, and refocuses the image with a camera onto a 9k x 9k scientific detector. The design also includes actuators for the AFC System.

\subsection{Active Flexure Compensation System}
\label{sec:afc}
\subsubsection{Description}
\label{sec:afc-description}
The Active Flexure Compensation (AFC) functionality is meant to address the following issues:

\begin{itemize}
\item
  Ensure the daytime calibration is valid to reduce night-time data. The accuracy required is 0.01 pixels along the spectral direction and 0.2 pixels along the spatial direction.
\item
  Avoid spectral resolution and efficiency loss due to mechanical flexures of the instrument during the exposure.
\end{itemize}

In order to achieve these goals, the procedure takes into account two reference spectra (one for each arm of the spectrograph) of the AFC lamp, taken with the telescope in the parking position, and associated to the latest wavelength calibration files. During observations, the rigid shifts of the AFC spectra in both spatial and spectral directions will be analyzed (Sect. \ref{sec:afc-algorithm}). When flexures due to telescope attitude are detected, the AFC actuators will be used to compensate their effect. We expect to acquire AFC spectra images continuously during science observations, i.e. the AFC lamp will always be ON during operations. The image of the AFC spectrum in this initial acquisition, as well as the image of the last AFC acquisition (if present) will be attached as an extension to the science FITS file for further analysis by the DRS (e.g. to correct for second order effects such as spectral ``breathing'').

\subsubsection{Cross-correlation Algorithm}
\label{sec:afc-algorithm}
The CUBES control software shall be able to estimate the shifts (in both spatial and spectral directions) of the AFC lamp spectrum as seen on both science detectors during observations.

First, a high SNR reference spectrum of the AFC lamp is taken during daytime calibrations. During night-time observations, new spectra with low SNR are taken at regular intervals. The region containing the spectrum (9200 x 60 pixels) is selected from both the observation and reference frames. Bias is subtracted, and the images are cross-correlated with a lag spanning 50 pixels in both directions, resulting in a 2D cross-correlation array. The peak of this array indicates the shift, and a 2D Gaussian fit is used to find the peak position with sub-pixel accuracy, providing the best estimate of the spectral and spatial shift.

\subsection{Results}
\label{sec:afc-results}

In order to test the accuracy of the algorithm in estimating the shifts, a set of simulations obtained with the CUBES End-to-end (E2E) simulator was used.
Specifically, reference and observation frames of the Active Flexure Compensation (AFC) were generated, considering the spectrum of a ThAr lamp using Photron lines as reference. Observation frames were generated with multiple rigid shifts in the spectral and spatial direction and an exposure time equal to  one-tenth of that of the reference frame. As a first simulation step and to aid the development of the cross-correlation algorithm, the frames were provided without overscan regions.

\begin{figure}[h!]
  \centering
  \includegraphics[width=0.9\textwidth]{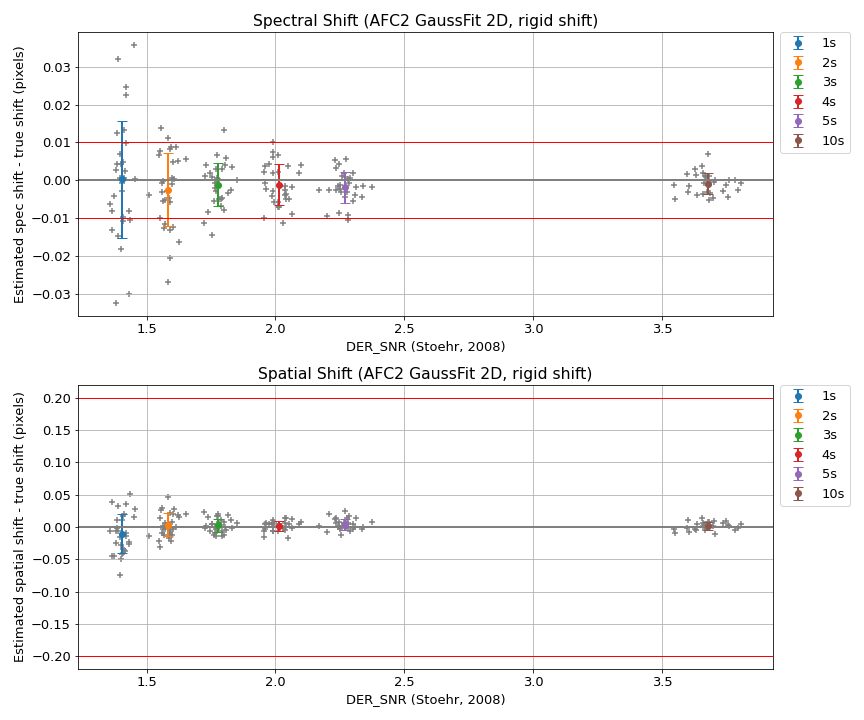}
  \caption{The result of the test of the AFC algorithm with simulated images. In the upper panel, the difference between the estimated spectral shift and the real value is shown, with each gray cross corresponding to a single image and the dots with error bars corresponding to the mean $\pm \sigma$. In the lower panel, the same plot for the spatial direction. Red lines show the required accuracy.}
  \label{fig:afc-test-p3}
\end{figure}

As shown in Figure \ref{fig:afc-test-p3}, the distribution of the best fit values for the center of the 2D model curve shows that the algorithm is able to estimate the shift with the required accuracy of 0.01 pixels along the spectral direction and 0.2 pixels along the spatial direction. Refer here \cite{cubes_2024,modeling_cubes_2024} for more information.

\begin{figure}[h!]
  \centering
  \includegraphics[width=0.9\textwidth]{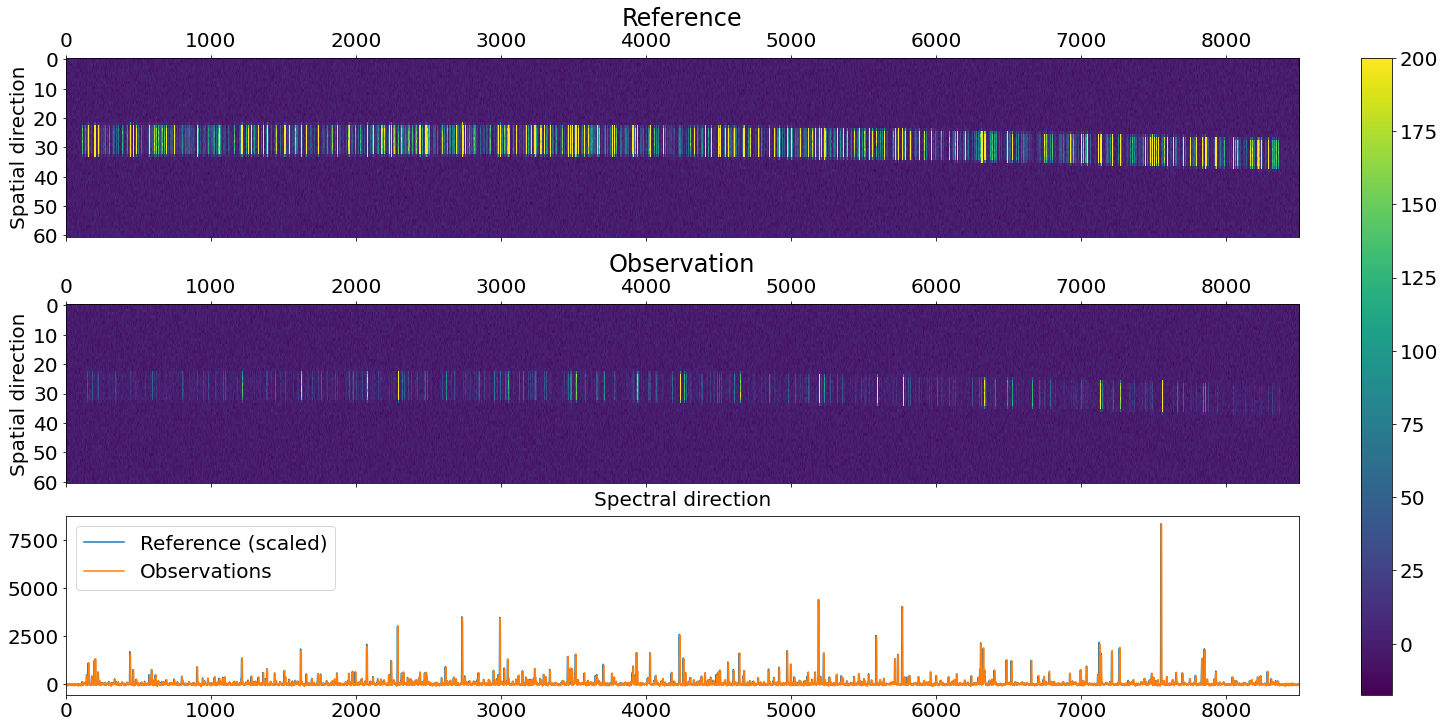}
  \caption{The daytime reference image (upper panel) and the night-time observation image (middle panel). In the lower panel, the reference and observation spectra are shown, obtained by summing each image along the spatial direction.}
  \label{fig:afc-obs-ref}
\end{figure}

\begin{figure}[h!]
  \centering
  \includegraphics[width=\textwidth]{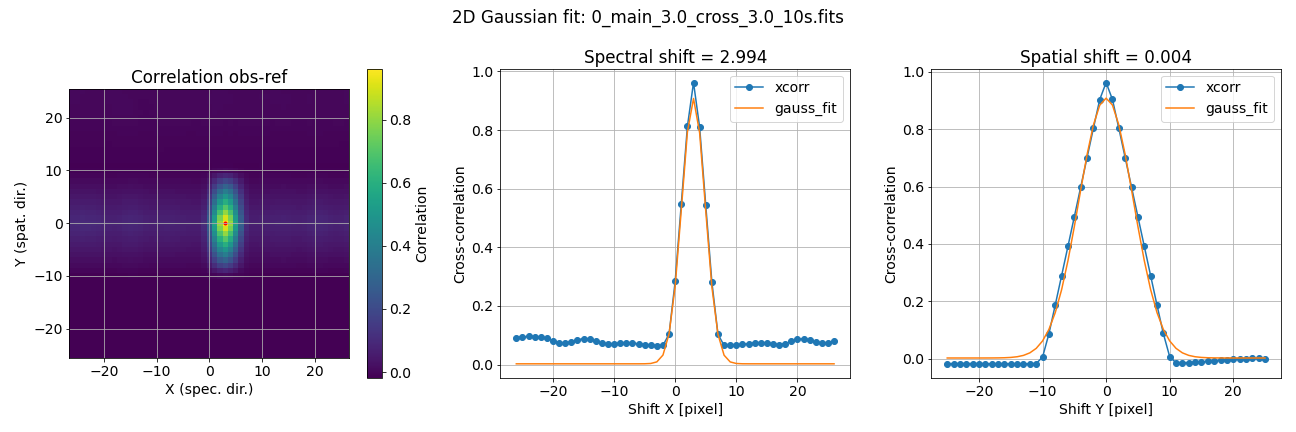}
  \caption{
  \textbf{Left panel}: the 2D cross-correlation between the observation and reference frames. 
  \textbf{Middle panel}: a horizontal slice of the 2D cross-correlation (in blue) and of the best-fit model (in orange), at the position of the center.
  \textbf{Right panel}: a vertical slice of the 2D cross-correlation (in blue) and of the best-fit model (in orange), at the position of the center.}
  \label{fig:afc-xcorr}
\end{figure}

\subsection{Data Reduction Software}
\begin{figure}[ht]
  \centering
  \includegraphics[width=0.9\textwidth]{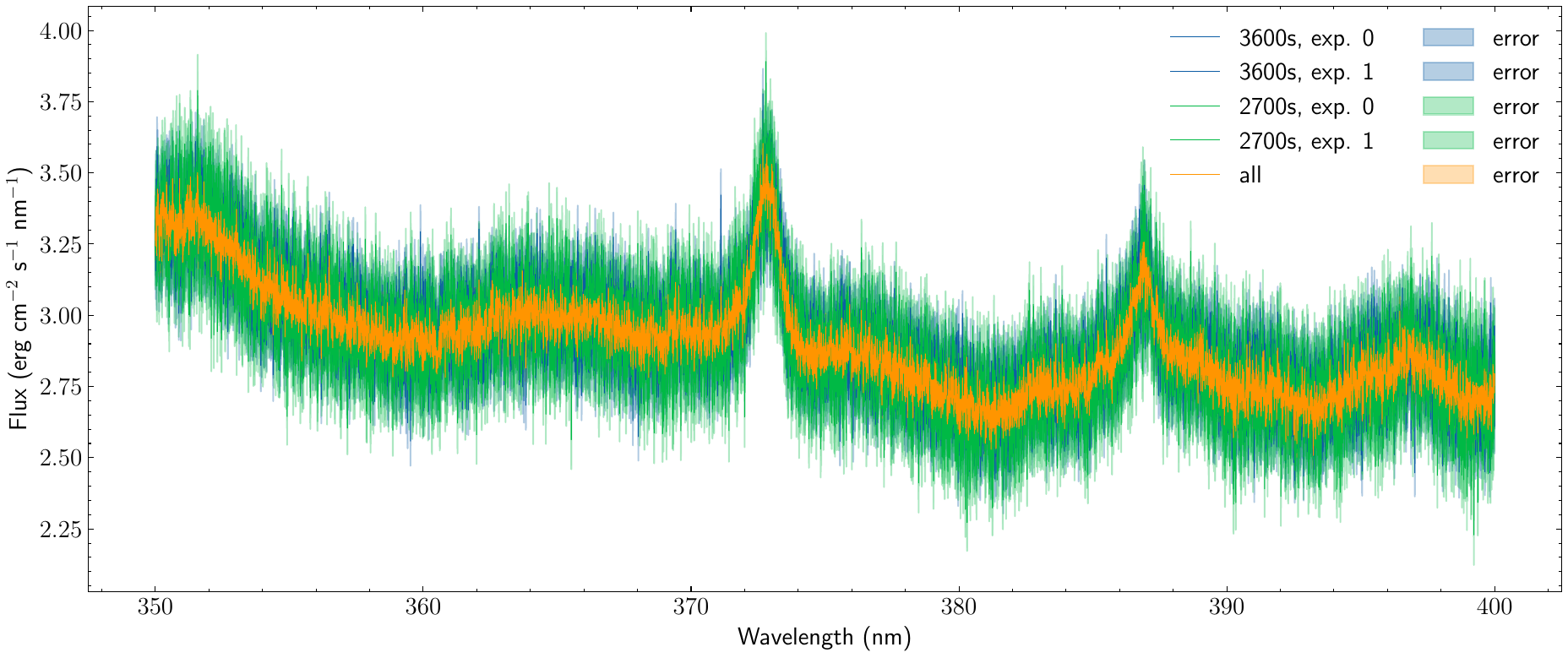}
  \caption{Extracted quasar spectra, produced by the prototype pipeline from E2E simulated data. The purpose is to compare the result of the extraction for individual exposures of the same target (blue and green; integration time of 3600 s and 2700 s respectively) with the co-added spectrum obtained by extracting all exposures at once (orange). The increase in SNR is apparent, as well as the preservation of the correct flux levels.}
  \label{fig:extr}
\end{figure}

\begin{figure}[ht]
  \centering
  \includegraphics[width=0.9\textwidth]{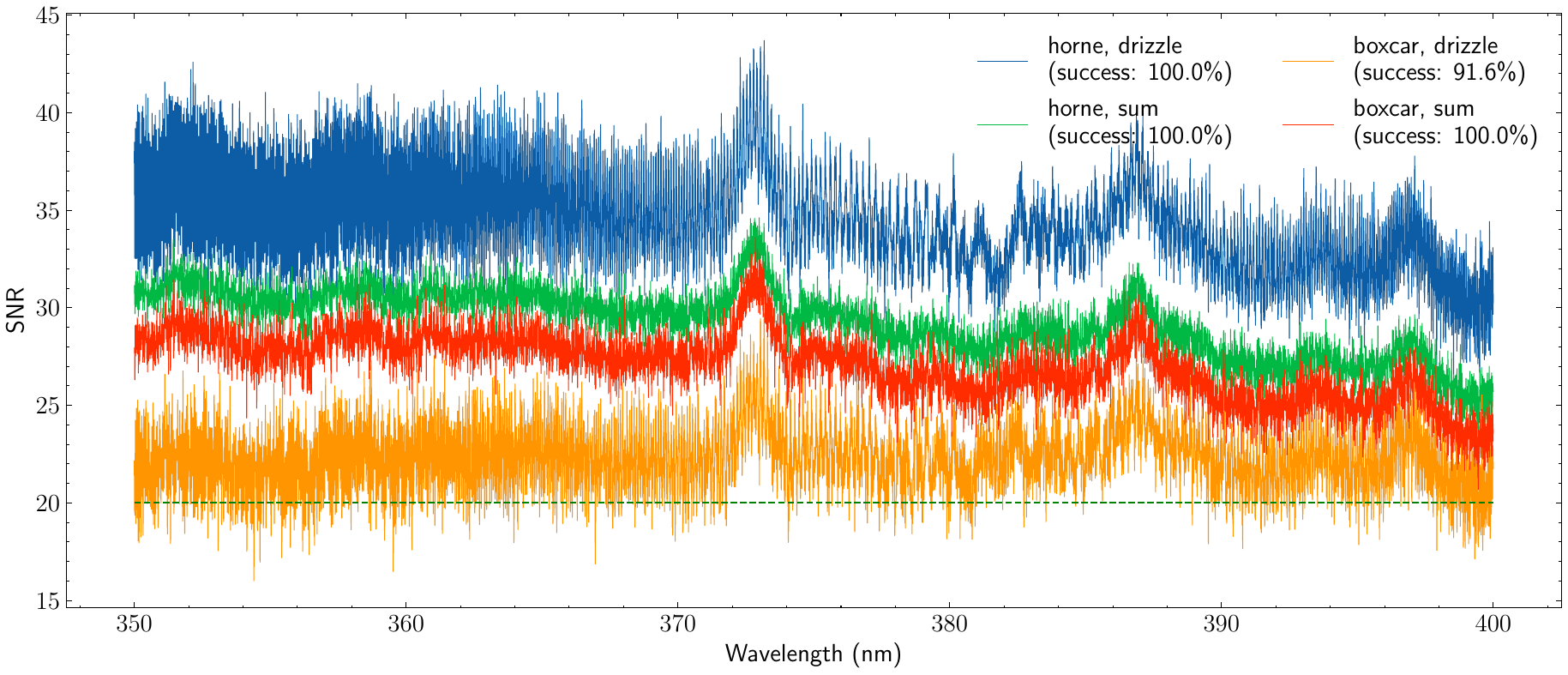}
  \caption{SNR obtained with different extraction techniques, namely (cf. text): (i) drizzling with Horne-like weighting (blue), (ii) sum with Horne-like weighting (green), (iii) drizzling with boxcar weighting (orange), and (iv) sum with boxcar weighting (red). The advantage of the first approach is apparent. The dashed line shows the threshold SNR to be achieved on a 18 magnitude target in 3600 s, according to the top-level requirements for the instrument.}
  \label{fig:snr}
\end{figure}

\begin{figure}[ht]
  \centering
  \includegraphics[width=0.9\textwidth]{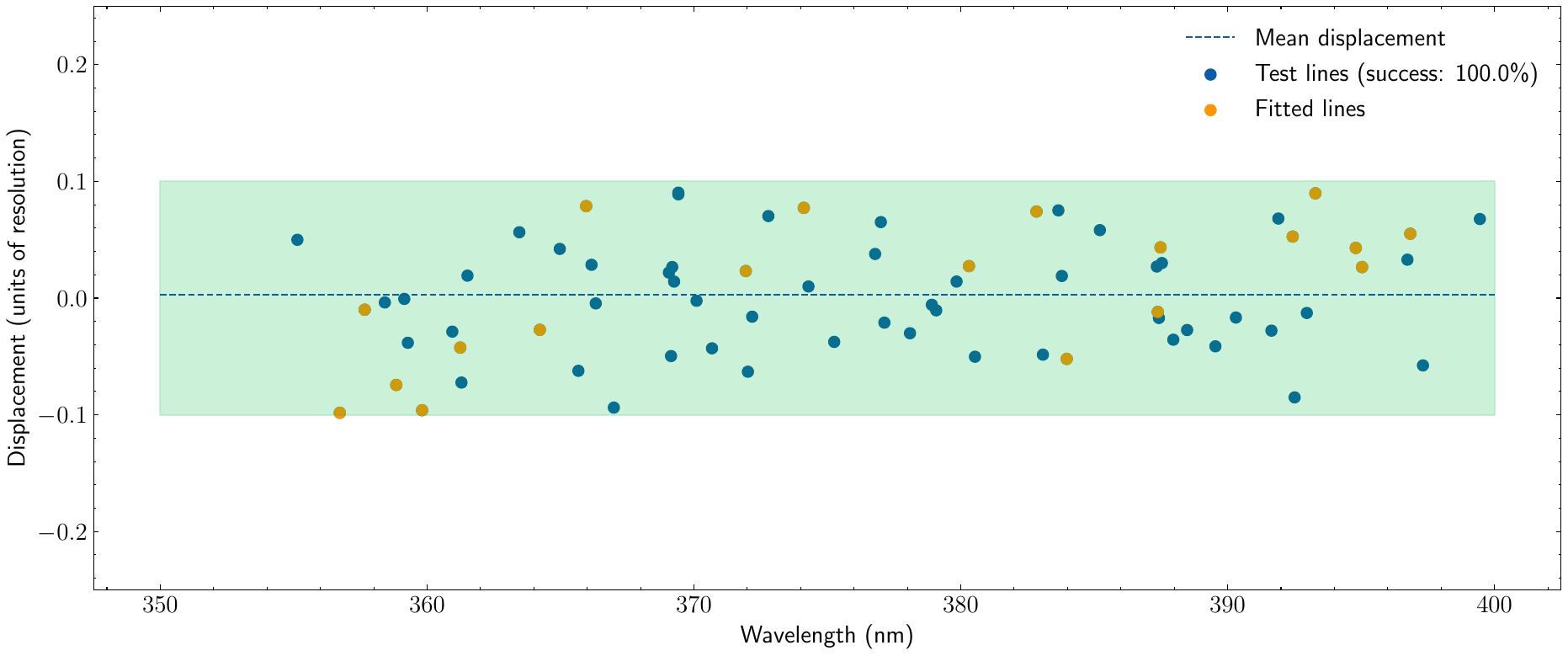}
  \caption{Displacements between laboratory wavelengths of selected ThAr lines and calibrated wavelengths measured on the extracted ThAr spectrum. Orange dots are the lines used to produce the 2-d wavelength map, while blue dots are other test lines used to assess the accuracy of the wavelength calibration. The top-level requirements for the instrument set an accuracy limit of 0.1 resolution elements, corresponding to the green shaded region.}
  \label{fig:wave}
\end{figure}

\begin{figure}[htp]
\begin{center}
\subfloat[Science frame: QSO object]{\label{fig:cubes_e2e_qso}\includegraphics[width=\linewidth]{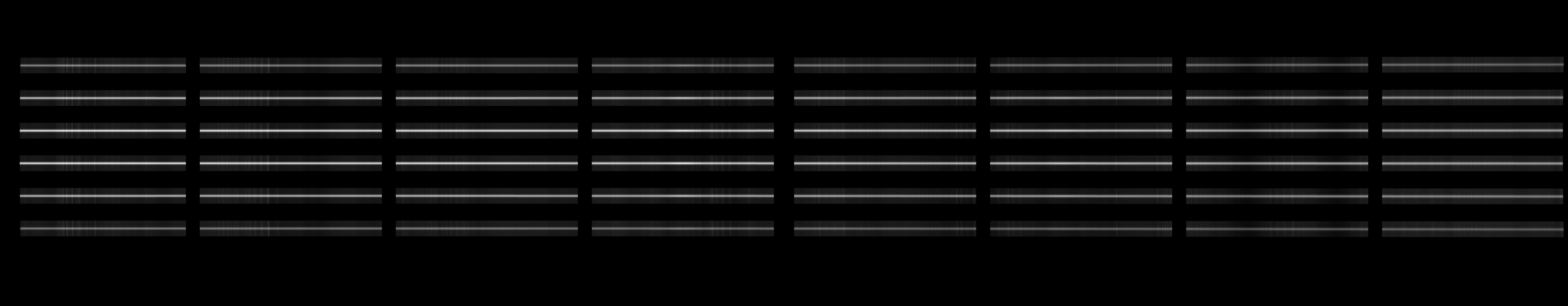}} \qquad
\subfloat[Calibration frame: Thorium-Argon (ThAr) lamp]
{\label{fig:cubes_e2e_calib}\includegraphics[width=\linewidth]{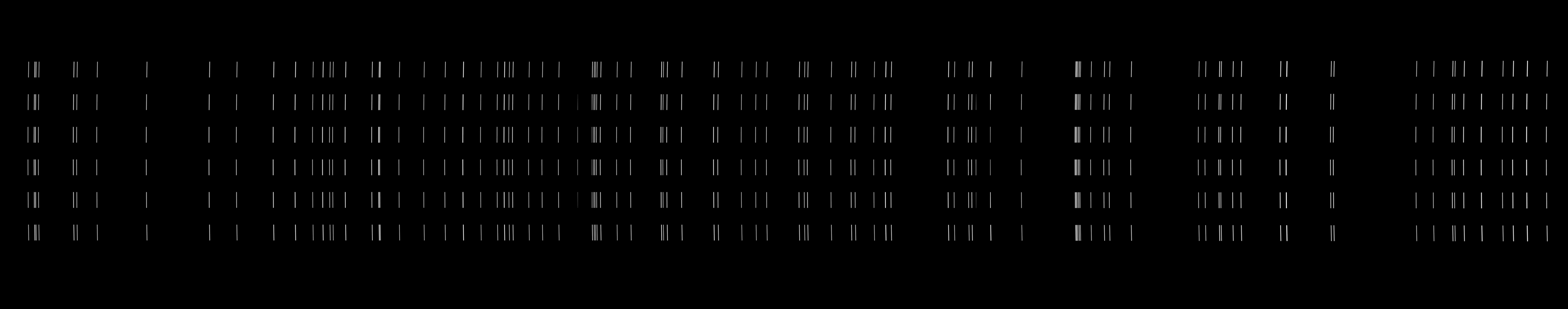}}
\caption[example] 
{ \label{fig:cubes_e2e_frames}  
Portions of raw simulated frames used in the development of optimal extraction and wavelength calibration recipes. The above image (\ref{fig:cubes_e2e_qso}) shows a raw science frame of a QSO object (which also includes sky contribution), while the bottom (\ref{fig:cubes_e2e_calib}) shows a raw calibration frame with a Thorium-Argon (ThAr) lamp. The detector noise in the original frames has been removed for better visualization.
}
\end{center}
\end{figure}

The CUBES Data Reduction Software (DRS) is now in the final design phase (see details here\cite{modeling_cubes_2024}). It includes three distinct pipelines:
\begin{itemize}
\item a pipeline to reduce and extract spectroscopic data, correcting from the detector signature, and performing wavelength calibration, sky subtraction, and standard-star-based flux calibration; 
\item a pipeline to perform photometry on the A\&G images, in order to improve the absolute flux calibration of the spectroscopic data; 
\item a pipeline to monitor the instrument health status, performing wavelength calibration and checks on the detector linearity.
\end{itemize}
\newpage
The core of the DRS consists of a set of ``recipes'' that implement the following critical algorithms:
\begin{itemize}
\item Wavelength calibration: a set of isolated, high-SNR ThAr lines are detected in a calibration spectrum and fitted in with a bivariate spline, to obtain the 2-d map between positions in the detector space and wavelengths. All pixels in the 6 slices are thus assigned a calibrated wavelength.
\item Sky modeling: the emission spectrum of the sky (continuum and lines) is sampled using the spatial extension of the slices\cite{2003PASP..115..688K}, either directly on the science target observation or on a dedicated sky observation, and smoothed with a B-spline.
\item Flux calibration: the response curve of the instrument is interpolated at the wavelength of all pixels in the slices, to convert electron counts in flux units. After extraction, flux calibration is further improved by using the photometry from the A\&G camera.
\item Extraction and co-addition: an optimal-extraction algorithm\cite{1986PASP...98..609H} is combined with a ``drizzling-like'' technique to weight the contributions to each wavelength bin in the final spectrum\cite{2016SPIE.9913E..1TC}, maximizing the SNR per resolution element while preserving flux calibration.
\end{itemize}
Already during Phase C, a complete prototype of the critical algorithms has been implemented (as a suite of Python scripts and Jupyter notebooks\cite{Kluyver2016jupyter}). Some products of the prototype pipeline are given in Figures \ref{fig:extr}, \ref{fig:snr}, and \ref{fig:wave}, showing respectively the result of extraction and co-addition of spectra, the SNR achieved by different extraction techniques, and the accuracy of wavelength calibration. These products are obtained from the simulated frames shown in Figure \ref{fig:cubes_e2e_frames}, and the results demonstrate how the prototype pipeline (and the full DRS when it is implemented) is an essential complement to the E2E simulator, effectively closing the loop from the input template spectra to the final extracted products. Throughout the integration phase, the joint E2E+DRS infrastructure will be pivotal in checking the performances of both the simulator and the reduction algorithms.

The final version of the DRS, to be released by ESO after the instrument commissioning, will be implemented using the ESO Common Pipeline Library (CPL)\footnote{\url{https://www.eso.org/sci/software/cpl/}.} and will be accesses through the new Python-based ESO Data Processing System (EDPS)\cite{2024A&A...681A..93F}. 


\label{sec:drs}

\newpage
\section{The ANDES Instrument}
\label{sec:andes_instrument}
\subsection{ANDES instrument overview}
\begin{figure} [h!]
\begin{center}
\begin{tabular}{c} 
\includegraphics[width=0.8\linewidth]{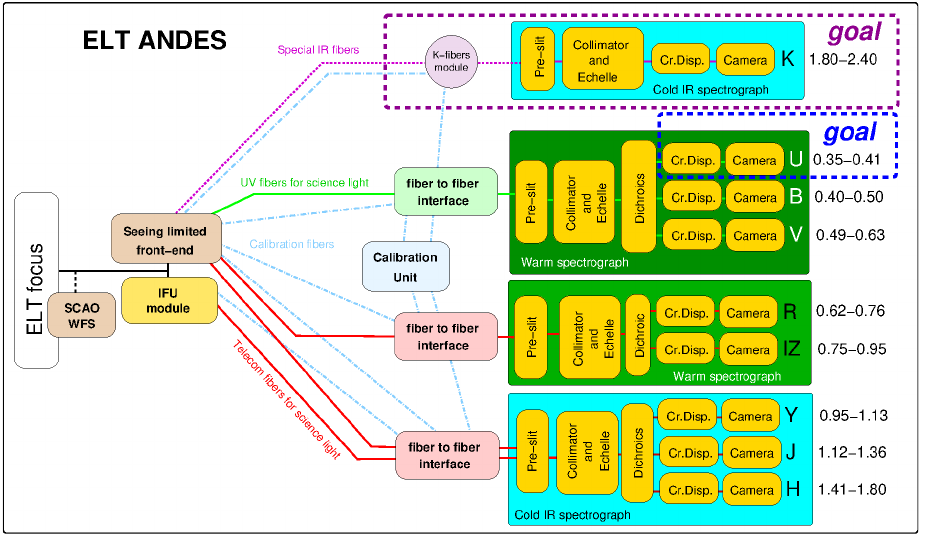}
\end{tabular}
\end{center}
\caption[example] 
{ \label{fig:ANDES_baseline} 
Reference design of ANDES and its modules}
\end{figure}

The baseline design for the ANDES spectrograph \cite{andes_2024} features a modular, fiber-fed, cross-dispersed echelle spectrograph with three ultra-stable spectrographs, each with several arms: (U)BV, RIZ, and YJH. This configuration provides a simultaneous spectral range of 0.4-1.8 $\mu$m at a resolution of approximately $\sim$100,000. The goal is to extend the wavelength range to 0.35-2.4 $\mu$m by adding a U extension and an additional spectrograph for K-band. The instrument will be able to operate both in a seeing-limited mode, and in a SCAO-assisted IFU-mode. In seeing-limited mode, simultaneous sky and/or calibration measurements are possible. The SCAO mode features a small IFU with at least two selectable spaxel scales (available in the YJH arm, with an extension to the RIZ arm as a goal). The proposed baseline of the system is illustrated in Fig. \ref{fig:ANDES_baseline}, which presents a schematic view of the functional architecture and highlights the chosen level of modularity for the instrument. Briefly, the main subsystems of ANDES are listed below.
\subsubsection{Front End}
The Front-End includes four modules: two seeing-limited arms, one SCAO arm, and one IFU arm. Depending on the chosen operational mode, the instrument can insert different modules and direct the light accordingly. Regardless of the specific observing mode, the Front-End splits the light from the telescope using dichroics, provides the necessary optical corrections, and delivers the light to the fiber bundles.

\subsubsection{Fiber Link}
The Fiber-Link subsystem is responsible for transporting light to the spectrographs and forming a series of parallel entrance slits. Each slit consists of an array of micro-lenses attached to a fiber bundle. The fiber bundles are divided into two segments, each following a different routing path. The first segment connects the FE to the Fibre-to-Fibre interface boxes, while the second segment connects the Fibre-to-Fibre interface boxes to the spectrographs.

\subsubsection{Calibration Unit (CU)}
The calibration unit includes proper calibration sources (HCL, Fabry-Perot, LFC) for feeding the Front-End and the Fibre-to-Fibre interfaces (part of the FL) through a suitable routing of fibres. The calibration Unit consists of three independent units which cover the bandwidth of the respected spectrographs. In detail, one CU will be placed on the Nasmyth platform for the BV spectrometer, while two other units (for the RIZ and YJH spectrometers) will be placed in the Coudé. If the K band is implemented, an additional CU unit will be placed on the Nasmyth platform.

\subsubsection{Single Conjugated Adaptive Optics (SCAO)} 
The SCAO aimed to provide the adaptive optics correction to the IFU feeding the YJH spectrograph and, in the case of its implementation, to the IFU feeding the K-band spectrograph. The SCAO system consists of a set of non-real-time hardware and a TCCD, which is designed to provide an atmospheric dispersion corrected beam to the pyramid and to ensure proper alignment and centering on the Wavefront Sensor (WFS). The WFS then feeds data to the Real-Time Computer, which calculates the necessary deformations to be applied to the telescope's M4 and M5 mirrors to compensate for atmospheric light distortion.

\subsection{Data Reduction Software}
\label{sec:andes_drs}

The ANDES DRS (more details can be found here \cite{andes_sw_2024}) is in an early planning phase. The team builds of the experience from current high resolution and precision spectrographs (HARPS, ESPRESSO, NIRPS) and also slit-spectrographs like CRIRES, since the ANDES will be tightly packed into a pseudo-slit. One of the crucial questions in seeing-limited mode is therefore, whether the overlapping fibers can be treated as a true slit, albeit with inclined and curved, for extraction in to 1D spectra. Conversely, in the IFU-mode the fibers will need to be treated individually to retain the spatial resolution, and therefore the cross-talk between fibers should be minimized.

\begin{figure}
    \centering
    \includegraphics[width=0.75\linewidth]{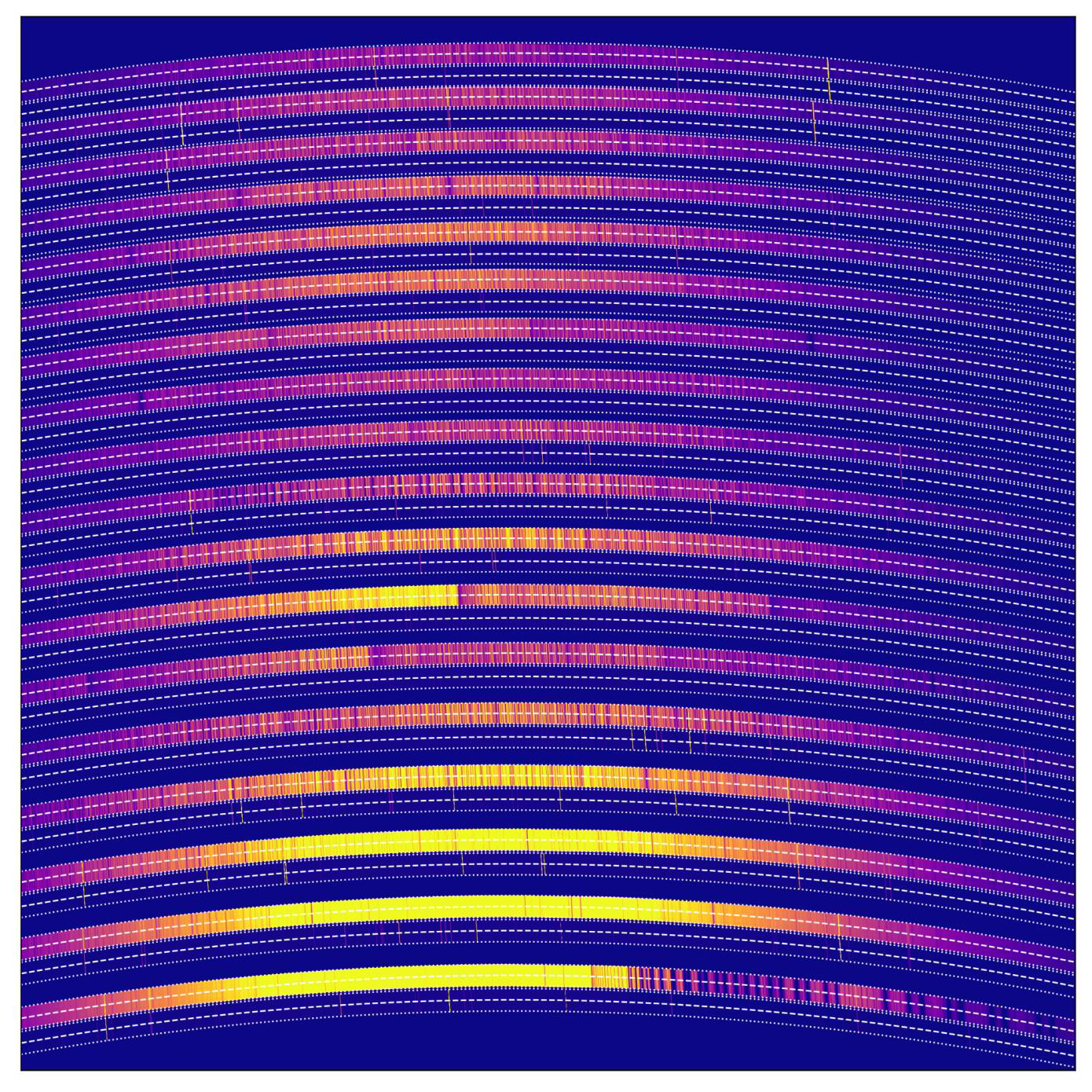}
    \caption{RIZ synthetic image (R band) obtained using the ANDES End to End simulator showing the dispersion of the two pseudo-slits on the detector with the calibration fibre in the middle. Plotted on top in white dashed and dotted lines are the mid-lines and edge of the slit, as determined by our DRS prototypes.}
    \label{fig:e2e-riz}
\end{figure}

We used the End to End simulator to gain better insight into the data format by synthesizing a raw frame secured by the RIZ spectrometer (R band) during seeing-limited target+sky observation with simultaneous wavelength calibrations. As shown in Figure \ref{fig:e2e-riz}, the RIZ synthetic image obtained using the ANDES End to End simulator displays the dispersion of the two pseudo-slits on the detector with the calibration fibers in the middle. In Aperture A, a G2V star spectrum with V = 10.00 is recorded simultaneously with the sky spectrum in Aperture B, and the simultaneous wavelength calibration is visible in the middle.

\begin{figure}
    \centering
    \includegraphics[width=0.8\linewidth]{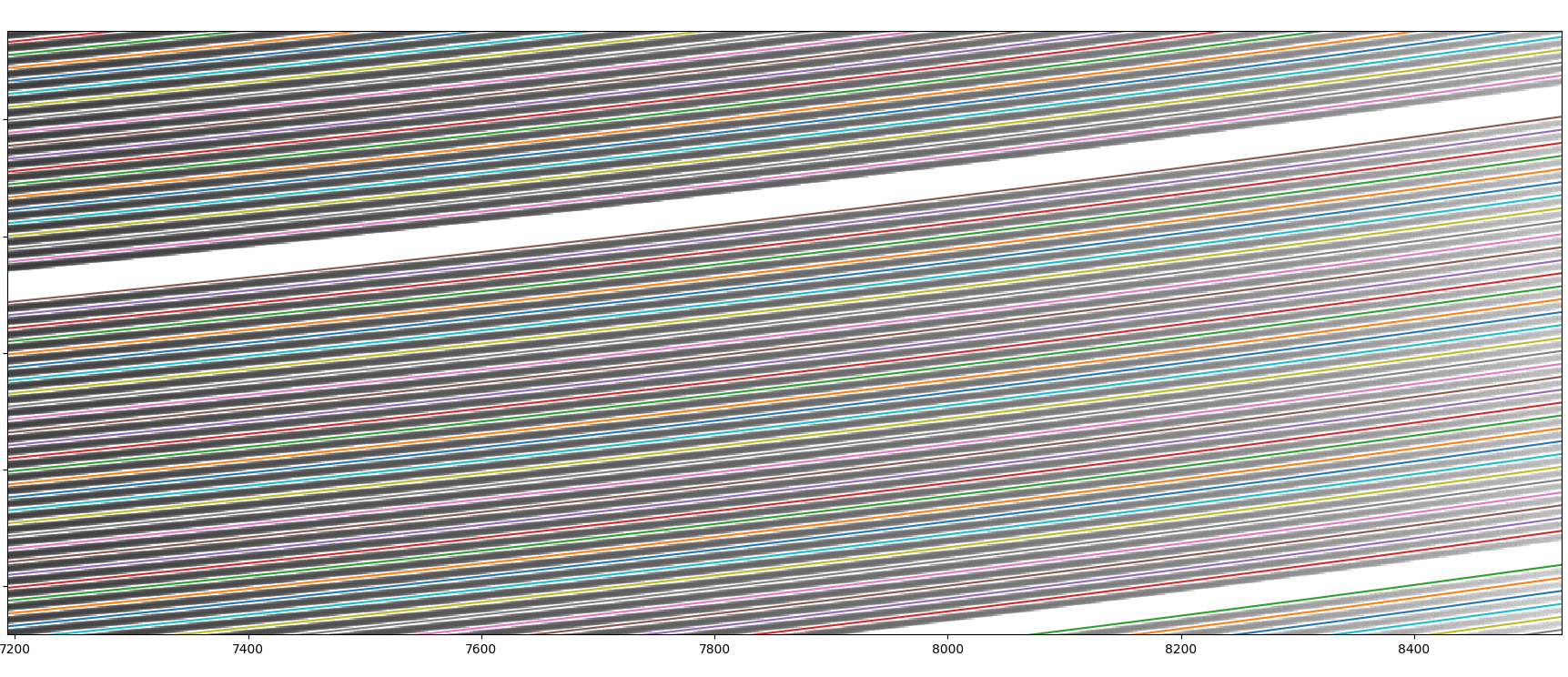}
    \caption{Zoom-in on a spectral order in the U-band. The grayscale image shows the E2E-simulation with only the odd-numbered fibers (33 out of 64) illuminated with a flat-field source. This frame and the corresponding one for the even fibers were used to trace the fiber locations in the frame. The resulting polynomials for the even fibers are overplotted as colored lines.}
    \label{fig:utrace}
\end{figure}

Using the E2E frames allows us to test algorithms and give illustrate how design decisions like fiber spacing and ordering impact the echellograms and thus the final spectra (see Figure \ref{fig:utrace}). The next step is to simulate the wavelength calibration sources and close the loop from template spectra to final reductions, and thereby assess the feasibility of ANDES' ambitious goals with respect to precision and accuracy.

\section{Conclusion}
We presented the End-to-End (E2E) model package applied to various astronomical spectrographs, including SOXS (ESO-NTT), CUBES (ESO-VLT), and ANDES (ESO-ELT). We described our simulation approaches, highlighting both our own E2E simulator model with its architecture and functionalities, and the external simulation frameworks, such as Pyechelle and Pyxel, utilized in our simulation process.
For each spectrograph, currently at different stages of development, we detailed how the generated high-fidelity synthetic frames have supported their respective phases. For the SOXS instrument, we demonstrated how the E2E model was employed during the Assembly, Integration, and Testing (AIT) phase. For the CUBES instrument, we discussed the work done to test the Active Flexure Compensation (AFC) cross-correlation algorithm and the contributions to developing the prototype pipeline. Finally, for the ANDES spectrograph, we outlined the iterations with the Data Reduction Software (DRS) team.

\label{sec:conclusion}
\bibliography{report} 

\begin{thebibliography}{10}

\bibitem{soxs_gen_2022_e2e}
Genoni, M., Scaudo, A., et~al., ``Progress on the simulation tools for the soxs spectrograph: Exposure time calculator and end-to-end simulator,'' (2022).

\bibitem{skycalc}
ESO, ``Skycalc,'' (2013).
\newblock \url{http://www.eso.org/observing/etc/bin/gen/form?INS.MODE=swspectr+INS.NAME=SKYCALC}.

\bibitem{numba}
Lam, S.~K., Pitrou, A., and Seibert, S., ``Numba: A llvm-based python jit compiler,'' in [{\em Proceedings of the Second Workshop on the LLVM Compiler Infrastructure in HPC}{\nolinebreak\hspace{0.1em}]},  {\em LLVM '15}, Association for Computing Machinery, New York, NY, USA (2015).

\bibitem{pyxel_2022}
Arko, M., Prod’homme, T., Lemmel, F., Serra, B., George, E.~M., Kelman, B., Pichon, T., Biancalani, E., and Gilbert, J., ``{Pyxel 1.0: an open source Python framework for detector and end-to-end instrument simulation},'' {\em Journal of Astronomical Telescopes, Instruments, and Systems}~{\bf 8}(4),  048002 (2022).

\bibitem{pyechelle_2019}
Stürmer, J., Seifahrt, A., Robertson, Z., Schwab, C., and Bean, J.~L., ``Echelle++, a fast generic spectrum simulator,'' {\em Publications of the Astronomical Society of the Pacific}~{\bf 131},  024502 (dec 2018).

\bibitem{soxs_update}
Schipani, P., Campana, S., Claudi, R., Aliverti, M., et~al., ``Development status of the {SOXS} spectrograph for the {ESO}-{NTT} telescope,'' in [{\em Ground-based and Airborne Instrumentation for Astronomy {VIII}}{\nolinebreak\hspace{0.1em}]},  Evans, C.~J., Bryant, J.~J., and Motohara, K., eds., {SPIE} (dec 2020).

\bibitem{soxs_gen_2022}
Schipani, P. et~al., ``Progress on the {SOXS} transients chaser for the {ESO}-{NTT},'' {\em Proc. SPIE} {\bf 12184-24} (2022).

\bibitem{soxs_cp}
Claudi, R. et~al., ``{The common path of SOXS (Son of X-Shooter)},'' {\em Proc. SPIE} {\bf 10702},  1189 -- 1199 (2018).

\bibitem{acq_guid_soxs}
Brucalassi, A. et~al., ``{Final Design and development status of the Acquisition and Guiding System for {SOXS}},'' {\em Proc. SPIE} {\bf 114475V} (2020).

\bibitem{MITS}
Rubin, A. et~al., ``Mits: the multi-imaging transient spectrograph for soxs,'' {\em Proc. SPIE} {\bf 10702} (2018).

\bibitem{soxs_vis}
Cosentino, R. et~al., ``The vis detector system of {SOXS},'' {\em Proc. SPIE} {\bf 10702} (2018).

\bibitem{soxs_optical_design}
S{\'{a}}nchez, R.~Z. et~al., ``Optical design of the {SOXS} spectrograph for {ESO} {NTT},'' {\em Proc. SPIE} {\bf 10702} (jul 2018).

\bibitem{soxs_nir_paper}
Vitali, F. et~al., ``{The NIR spectrograph for the new SOXS instrument at the NTT},'' {\em Proc. SPIE} {\bf 10702},  697 -- 709 (2018).

\bibitem{soxs_cal}
Kuncarayakti, H. et~al., ``Design and development of the {SOXS} calibration unit,'' {\em Proc. SPIE} {\bf 11447} (dec 2020).

\bibitem{soxs_gen_2024}
Genoni, M. et~al., ``Soxs nir: optomechanical integration and alignment, optical performance verification before full instrument assembly,'' {\em Proc. SPIE} {\bf 13096-104} (2024).

\bibitem{soxs_cp_2022}
Radhakrishnan, K. et~al., ``From assembly to the complete integration and verification of the {SOXS} common path,'' {\em Proc. SPIE} {\bf 12184-305} (2022).

\bibitem{soxs_acq_2022}
Araiza-Dur{\'{a}}n, J.~A. et~al., ``The integration and alignment phase for the acquisition and guiding system of {SOXS},'' {\em Proc. SPIE} {\bf 12184-306} (2022).

\bibitem{soxs_nir_2022}
Vitali, F. et~al., ``Progress on the {SOXS} {NIR} spectrograph {AIT},'' {\em Proc. SPIE} {\bf 12184-302} (2022).

\bibitem{cubes_2024}
Genoni, M. et~al., ``Cubes, the cassegrain u-band efficient spectrograph: Towards final design review.,'' {\em Proc. SPIE} {\bf 13096-296} (2024).

\bibitem{modeling_cubes_2024}
Scaudo, A. et~al., ``Modeling cubes: from instrument simulation to data reduction prototype.,'' {\em Proc. SPIE} {\bf 13099-86} (2024).

\bibitem{2003PASP..115..688K}
{Kelson}, D.~D., ``{Optimal Techniques in Two-dimensional Spectroscopy: Background Subtraction for the 21st Century},'' {\em PASP}~{\bf 115}(808),  688--699 (2003).

\bibitem{1986PASP...98..609H}
{Horne}, K., ``{An optimal extraction algorithm for CCD spectroscopy.},'' {\em PASP}~{\bf 98},  609--617 (1986).

\bibitem{2016SPIE.9913E..1TC}
{Cupani}, G., {D'Odorico}, V., {Cristiani}, S., et~al., ``{Integrated data analysis in the age of precision spectroscopy: the ESPRESSO case},'' {\em Society of Photo-Optical Instrumentation Engineers (SPIE) Conference Series} {\bf 9913},  99131T (2016).

\bibitem{Kluyver2016jupyter}
Kluyver, T., Ragan-Kelley, B., P{\'e}rez, F., et~al., ``Jupyter notebooks -- a publishing format for reproducible computational workflows,''  87 -- 90, IOS Press (2016).

\bibitem{2024A&A...681A..93F}
{Freudling}, W., {Zampieri}, S., {Coccato}, L., et~al., ``{Adaptive data reduction workflows for astronomy: The ESO Data Processing System (EDPS)},'' {\em AAP}~{\bf 681},  A93 (2024).

\bibitem{andes_2024}
Marconi, A. et~al., ``Andes, the high resolution spectrograph for the elt: science goals, project overview and future developments,'' {\em Proc. SPIE} {\bf 13096-39} (2024).

\bibitem{andes_sw_2024}
Landoni, M. et~al., ``The armazones high dispersion echelle spectrograph (andes) software ecosystem,'' {\em Proc. SPIE} {\bf 13101-21} (2024).

\end{thebibliography}
\bibliographystyle{spiebib} 
\end{document}